\documentclass[aps, preprint,showpacs,superscriptaddress,groupedaddress]{revtex4-1}
\RequirePackage{fix-cm}
\usepackage{amsmath}
\usepackage{amssymb}
\usepackage{braket}
\usepackage{bm}
\begin{document}

\title{Richardson-Gaudin geminal wavefunctions in a Slater determinant basis
}


\author{Charles-\'Emile Fecteau}
\author{Fr\'ed\'eric Berthiaume}
\author{Meriem Khalfoun}
\author{Paul Andrew Johnson}
\email{paul.johnson@chm.ulaval.ca}

\affiliation{D\'{e}partement de chimie, Universit\'{e} Laval, Qu\'{e}bec, Qu\'{e}bec, Canada}

\begin{abstract}
Geminal wavefunctions have been employed to model strongly-correlated electrons. These wavefunctions represent products of weakly-correlated pairs of electrons and reasonable approximations are computable with polynomial cost. In particular, Richardson-Gaudin states have recently been employed as a variational ansatz. This contribution serves to explain the Richardson-Gaudin wavefunctions in the conventional language of quantum chemistry. 
\end{abstract}

\maketitle

\section{Introduction}
\label{intro}
Strong electron correlation remains an unsolved problem in quantum chemistry. In such systems the orbital picture breaks down and thus methods built upon a mean-field of electrons such as Hartree-Fock (HF) or Kohn-Sham Density Functional Theory are not sufficiently accurate. In such cases, it is more productive to employ wavefunctions built from weakly-interacting pairs of electrons, i.e. geminals. Geminals were considered early in quantum chemistry \cite{hurley:1953,silver:1969,silver:1970}, though long before the availability of commercial software. Recently, there has been a strong renewed interest in these types of wavefunctions \cite{coleman:1997,surjan:1999,kobayashi:2010,surjan:2012,neuscamman:2012,peter:2013,johnson:2013,stein:2014,boguslawski:2014a,boguslawski:2014b,boguslawski:2014c,tecmer:2014,henderson:2014a,henderson:2014b,shepherd:2014,bulik:2015,pastorczak:2015}. In particular, the antisymmetric product of 1-reference orbital geminals (AP1roG), or equivalently pair coupled-cluster doubles (pCCD), has been found to describe the energetics of many bond-breaking processes quite well. 

Recently we have employed another geminal wavefunction, the Richardson-Gaudin (RG) states \cite{richardson:1963,richardson:1964,richardson:1965,gaudin:1976}, as a variational ansatz for strongly-correlated electrons \cite{johnson:2020}. The results were promising though there remain issues to be addressed before it could be considered a black-box approach. These wavefunctions are also being employed in nuclear structure theory \cite{stijn:2017} as well as condensed matter physics \cite{claeys:2017a}. The advantage of RG states is that they are eigenvectors of model Hamiltonians. See refs \cite{dukelsky:2004,ortiz:2005} for a large variety of model Hamiltonians possible. Thus we have not a single wavefunction, but a complete set with which to construct perturbation theories and Green's functions in an analogous fashion to many-body theories built upon Hartree-Fock. Indeed this is our intention. The intermediate developments are tricky, but the final expressions are simple \cite{fecteau:2020}. Along similar lines, the group of Scuseria has built a mean-field theory based on the antisymmetrized geminal power (AGP) \cite{henderson:2019,khamoshi:2019,henderson:2020a,dutta:2020,harsha:2020,khamoshi:2020}. 

All of these geminal wavefunctions are approximations of the antisymmetrized product of interacting geminals (APIG) \cite{silver:1969,silver:1970}. APIG is the most general mean-field for pairs, and is thus intractable to compute with. For systems with only a few important pairs, AP1roG / pCCD is an excellent approximation to APIG, while for systems entirely dominated by pairs, RG is a better approximation. Both AP1roG / pCCD and RG scale polynomially (specifically $\mathcal{O}(M^4)$) whereas computations with APIG do not.

The purpose of this contribution is to understand the RG wavefunction in terms of Slater determinants, in particular its relationship with pair coupled-cluster wavefunctions. We will demonstrate that the cumulants have a simple closed form, from which cluster amplitudes may be reverse-engineered. Finally, while AP1roG / pCCD is the optimal first approximation to APIG in the Slater determinant basis, we add optimal second approximations as well as hierarchical approximations to higher orders.

\section{RG in Slater determinant basis}
\label{sec:1}
\subsection{APIG and RG}
With the usual electron creation and annihilation operators, we define the objects:
\begin{align} \label{eq:su2}
S^+_i &= a^{\dagger}_{i\uparrow}a^{\dagger}_{i\downarrow} \\
S^-_i &= a_{i\downarrow} a_{i\uparrow}
\end{align}
which have the action of creating or removing a pair of electrons from the spatial orbital $i$. Each spatial orbital can only contain one pair of electrons, so the repeated action of any individual pair creator is zero. With these pair creators, the most general  mean-field wavefunction in terms of the geminals
\begin{align}
G^+_{\alpha} = \sum_i g^i_{\alpha} S^+_i 
\end{align}
is the antisymmetric product of interacting geminals (APIG):
\begin{align}
\label{eq:APIG}
\ket{\text{APIG}} = G^+_1 G^+_2 \dots G^+_M \ket{\theta}
\end{align}
In eq. \eqref{eq:APIG}, $\ket{\theta}$ is a vacuum with respect to all the pair removal operators. APIG is a mean-field of geminals. With this wavefunction, one could in principle develop a variational theory like HF, though it would be intractable to compute with. The reason for this is simple. Expressed in a basis of Slater determinants, 
\begin{align}
\ket{  \text{APIG}} = \sum_{ \{i\}} C_{\{i\}} \ket{ \{i\}}
\end{align}
with $\{i\}$ denoting a collection of doubly-occupied spatial orbitals, and $\ket{\{i\}}$ the Slater determinant corresponding to this set, the expansion coefficients are symmetric sums, i.e. permanents, of the geminal coefficients
\begin{align} \label{eq:permanent}
C_{\{i\}} &= \sum_{\sigma} \prod_{\alpha=1}^{M} g^{i_{\alpha}}_{\sigma(\alpha)} 
= \text{per}\; \begin{pmatrix}
g^{i_1}_1 	& 	g^{i_1}_2 & \dots & g^{i_1}_M \\
g^{i_2}_1 	& g^{i_2}_2 & \dots & g^{i_2}_M \\
\vdots & \vdots & \ddots & \vdots \\
g^{i_M}_1 & g^{i_M}_2 & \dots & g^{i_M}_M
\end{pmatrix}.
\end{align}
Each geminal coefficient can only appear once, and there are no negative signs present as each permutation exchanges pairs of electrons. Permanents of matrices are intractable to compute in general \cite{minc:1978}. As determinants are invariants of matrices, they may be evaluated in $\mathcal{O}(M^3)$ operations by first diagonalizing the matrix. Permanents are not invariants of matrices, and hence must be calculated either by directly evaluating all $M!$ elements of the sum or by more effective, yet exponential, algorithms \cite{ryser:1963}. 

As written, APIG privileges no specific Slater determinant. In chemical systems, it is often the case that HF is a reasonable first approximation to APIG. In these cases, we can divide through by the coefficient of the HF Slater determinant and group the other terms as pair excitations of HF.
\begin{align} \label{eq:APIG_det}
\frac{1}{C_{HF}} \ket{ \text{APIG}} = \ket{\text{HF}} + \sum_{\substack { i \in occ \\ a \in virt}} C^a_i S^+_a S^-_i \ket{\text{HF}} 
+  \sum_{\substack { i,j \in occ \\ a,b \in virt}} C^{ab}_{ij} S^+_a S^+_b S^-_j S^-_i \ket{\text{HF}} + \dots
\end{align}
It is possible, though intractable, to reverse engineer this expression into an exponential form, that is to say a cluster operator. There is no further simplification for APIG. To proceed there are two approaches. The first is to choose $C_{HF}=1$ and treat the single and double excitations as variables, neglecting the rest. This is the choice made for AP1roG / pCCD:
\begin{subequations} \label{eq:ap1rog}
\begin{align} 
\ket{\text{AP1roG}} &= \prod_{ i \in occ} \left( S^+_i + \sum_{a \in virt} t^a_i S^+_a \right) \ket{\theta} \\
\ket{\text{pCCD}} &= \exp \left( \sum_{\substack{ i \in occ \\ a \in virt}} t^a_i S^+_a S^-_i \right) \ket{\text{HF}}
\end{align}
\end{subequations}
The other approach is to choose a structure for the geminal coefficients such that the permanents may be computed effectively. An example of this case is to choose the form of a Cauchy matrix,
\begin{align}
\label{eq:cauchy_elements}
g^i_{\alpha} = \frac{1}{u_{\alpha} - \varepsilon_i}
\end{align}
in terms of two sets of variables $\{\varepsilon\},\{u\}$. This is the choice for the RG wavefunction. With the local pair operators \eqref{eq:su2}, we define the geminals:

\begin{align}
S^+ (u) &= \sum_i \frac{S^+_i}{u - \varepsilon_i}
\end{align}
in terms of a complex number $u$. The RG wavefunction is the product of geminals
\begin{align}
\label{eq:rg}
\ket{ \text{RG} } = S^+ (u_1) S^+(u_2) \dots S^+(u_M) \ket{\theta}.
\end{align}
The state \eqref{eq:rg} is an eigenvector of the reduced Bardeen-Cooper-Schrieffer (BCS) Hamiltonian \cite{bardeen:1957a,bardeen:1957b}
\begin{align}
\label{eq:bcs_ham}
\hat{H}_{BCS} = \frac{1}{2} \sum_i \varepsilon_i \left( a^{\dagger}_{i \uparrow} a_{i\uparrow} + a^{\dagger}_{i \downarrow} a_{i\downarrow} \right) 
- \frac{g}{2} \sum_{ij} a^{\dagger}_{i\uparrow}a^{\dagger}_{i\downarrow} a_{j \downarrow} a_{j\uparrow}
\end{align}
provided that the complex numbers $\{u\}$ satisfy the set of coupled non-linear equations (Richardson's equations):
\begin{align}
0 = \frac{2}{g} + \sum_i \frac{1}{u_a- \varepsilon_i} + \sum_{b \neq a} \frac{2}{u_b - u_a},\quad \forall a=1,\dots M.
\end{align}
The reduced BCS Hamiltonian expresses competition between an aufbau filling of the lowest spatial orbitals, with energies $\{\varepsilon\}$, and an isotropic pairing interaction $g$.  

For the RG wavefunction, a variational approach is possible, and indeed feasible due to (the limiting case of) a result of Slavnov for its scalar products \cite{slavnov:1989,belliard:2019,zhou:2002}. Expressions for the reduced density matrices are computable with a small number of determinants or even more effectively as solutions of linear equations \cite{faribault:2008,faribault:2010,GB:2011,claeys:2017b,fecteau:2020}. The point of this contribution is to write the RG wavefunction in the basis of Hartree-Fock Slater determinants, ideally as the action of a coupled-cluster operator. As we will see, expressions for the cumulants are quite clean, while the cluster amplitudes are not (though they exist in any case). For the variational approach, it is absolutely essential that the complex numbers $\{u\}$ are solutions of Richardson's equations. In this contribution however, all of the results rely only on the structure of the geminal coefficients \eqref{eq:cauchy_elements} so we may take $\{u\}$ to be any complex numbers.

\subsection{RG: Cumulants}
To proceed, there are three tools we will need. The first is Borchardt's theorem \cite{borchardt}, which allows us to compute the permanent of a Cauchy matrix:
\begin{align}
\underset{i,\alpha}{\det}\left( \frac{1}{u_{\alpha}- \varepsilon_i} \right) \underset{i,\alpha}{\text{per}}\left( \frac{1}{u_{\alpha}- \varepsilon_i} \right) = \underset{i,\alpha}{\det}\left( \frac{1}{(u_{\alpha}- \varepsilon_i)^2} \right) \label{eq:borc}.
\end{align}

The second tool is Cramer's rule, which states that the solution of a system of linear equations
\begin{align}
Ax = b
\end{align}
has elements
\begin{align}
x_i  = \frac{\det(A^b_i)}{\det(A)}
\end{align}
where the matrix $A^b_i$ is the matrix $A$ with the $ith$ column replaced with the RHS $b$. 

The third tool is an identity of Jacobi relating determinants of matrices with multiple columns replaced \cite{vein_book}. For a matrix with two replaced columns,
\begin{align}
\frac{ \det (A^{ab}_{ij}) }{\det(A)} = \frac{  \det(A^a_i ) }{\det(A)} \frac{  \det(A^b_j ) }{\det(A)} - \frac{  \det(A^b_i ) }{\det(A)} \frac{  \det(A^a_j ) }{\det(A)}
\end{align}
where $A^{ab}_{ij}$ is the matrix $A$ with the $ith$ column replaced with the vector $a$ and the $jth$ column replaced with the vector $b$. The result is that the determinant of the matrix with two replaced columns, scaled by the determinant of the original matrix, is the determinant of the single replacements. Remarkably, this identity extends to any order. Specifically, the determinant of a matrix with $k$ replaced columns, scaled by the original determinant, is a $k\times k$ determinant of single column replacements.

As the RG state is a specific case of APIG, it can be written in a basis of Slater determinants in the same manner. Specifically, 
\begin{align}
\ket{\text{RG}} = \sum_{ \{i\}} C_{\{i\}} \ket{ \{i\}} \label{eq:rg_slat}.
\end{align}
with the expansion coefficients:
\begin{align}
C_{\{i\}} &= \sum_{\sigma} \prod_{\alpha=1}^{M}  \frac{1}{u_{\alpha} - \varepsilon_{\sigma (i_{\alpha})}}
= \underset{i,\alpha}{\text{per}}\left( \frac{1}{u_{\alpha}- \varepsilon_i} \right).
\end{align}
In this particular case, Borchardt's theorem allows us to calculate this permanent in terms of two determinants.

Again, in equation \eqref{eq:rg_slat} each Slater determinant is on equal footing. By dividing through by the expansion coefficient of the HF Slater determinant, we can regroup the terms in the expression based on the number of pair excitations on top of HF. Specifically, 
\begin{align} \label{eq:RGCC}
\frac{1}{C_{HF}} \ket{\text{RG}} &= \ket{\text{HF}} + \sum_{\substack { i \in occ \\ a \in virt}} C^a_i S^+_a S^-_i \ket{\text{HF}} 
+  \sum_{\substack { i,j \in occ \\ a,b \in virt}} C^{ab}_{ij} S^+_a S^+_b S^-_j S^-_i \ket{\text{HF}} + \dots \\
&= \left( \hat{1} + \hat{C}_1 + \hat{C}_2 + \dots + \hat{C}_M \right) \ket{\text{HF}}. \label{eq:RGCC2}
\end{align}
This expansion has the form of a paired coupled-cluster wavefunction written in terms of its cumulants rather than the usual exponential form. The cumulants have a particularly simple form as we will show.

The HF coefficient is the permanent of a Cauchy matrix, $B$, corresponding to putting pairs in the lowest energy orbitals
\begin{align}
C_{HF} = \text{per} (B),
\end{align}
with
\begin{align}
B = \begin{pmatrix}
\frac{1}{u_1 - \varepsilon_1} & \dots & \frac{1}{u_1 - \varepsilon_M} \\
 & \ddots \\
\frac{1}{u_M - \varepsilon_1} & \dots & \frac{1}{u_M - \varepsilon_M}
\end{pmatrix}.
\end{align}
Next, the coefficients of the single pair excitations are:
\begin{align}
C^a_i = \frac{\text{per}(B^a_i)}{\text{per}(B)}
\end{align}
where the matrix $B^a_i$ is the matrix $B$ with its $i$th column replaced with the $a$th column
\begin{align} \label{eq:RHS}
b^a = \begin{pmatrix}
\frac{1}{u_1 - \varepsilon_a} \\
\frac{1}{u_2 - \varepsilon_a} \\
\vdots \\
\frac{1}{u_M - \varepsilon_a}
\end{pmatrix},
\end{align}
that is to say the matrix $B$ with one occupied column replaced by one virtual column. Because $B$ and $B^a_i$ are both Cauchy matrices, Borchardt's theorem can be used to simplify this to
\begin{align}
C^a_i = \frac{\det (B^a_i * B^a_i)}{\det (B^a_i)} \frac{ \det (B)}{\det(B*B) }
\end{align}
where $*$ represents the Hadamard, or element-wise, matrix product: the elements of $B*B$ are the squares of the elements of $B$ etc.

Now, with Cramer's rule, the solutions of the linear equations:
\begin{align}
(B*B) X^a &= b^a*b^a \\
B Y^a &= b^a
\end{align}
are explicitly the ratios of the determinants
\begin{align} \label{eq:primitives}
X^a_i &= \frac{\det (B^a_i * B^a_i)}{\det(B*B)} \\
Y^a_i &= \frac{\det(B^a_i)}{\det (B)}
\end{align}
so that
\begin{align}
C^a_i = \frac{X^a_i}{Y^a_i}.
\end{align}
The first-order cumulant is thus
\begin{align}
\hat{C}_1 = \sum_{ \substack{i \in occ \\ a \in virt } } \frac{X^a_i}{Y^a_i} S^+_aS^-_i.
\end{align}

The same reasoning leads to the expression for the second-order cumulant
\begin{align}
\hat{C}_2 = \frac{1}{4} \sum_{\substack{i,j \in occ \\ a,b \in virt  }} \frac{X^{ab}_{ij}}{Y^{ab}_{ij}} S^+_a S^+_b S^-_j S^-_i
\end{align}
with the notation
\begin{align}
X^{ab}_{ij} &= \frac{\det (B^{ab}_{ij} * B^{ab}_{ij})}{\det(B*B)} \\
Y^{ab}_{ij} &= \frac{\det(B^{ab}_{ij})}{\det (B)}
\end{align}
where the matrix $B^{ab}_{ij}$ is the matrix $B$ with the $i$th column replaced with the $a$th version of eq \eqref{eq:RHS} and the $j$th column replaced with the $b$th version of \eqref{eq:RHS}. Now, Jacobi's theorem reduces this result to the primitives \eqref{eq:primitives} already computed:
\begin{align} \label{eq:reduction}
X^{ab}_{ij} &= X^a_i X^b_j - X^a_j X^b_i \\
Y^{ab}_{ij} &= Y^a_i Y^b_j - Y^a_j Y^b_i.
\end{align}
Thus the second-order cumulant may be computed easily with the same information required for the first-order cumulant.

This extends to any order. Specifically,
\begin{align} \label{eq:big_cumulant}
\hat{C}_M = \left( \frac{1}{M!} \right) ^2 \sum_{\substack{i_1 \dots i_M \in occ \\ a_1\dots a_M \in virt}} \frac{X^{a_1 \dots a_M}_{i_1\dots i_M}}{Y^{a_1 \dots a_M}_{i_1\dots i_M}} S^+_{a_1}\dots S^+_{a_M}S^-_{i_M}\dots S^-_{i_1}
\end{align}
where 
\begin{align}
X^{a_1 \dots a_M}_{i_1\dots i_M} &= \frac{\det (B^{a_1 \dots a_M}_{i_1\dots i_M} * B^{a_1 \dots a_M}_{i_1\dots i_M})}{\det(B*B)} \\
Y^{a_1 \dots a_M}_{i_1\dots i_M} &= \frac{\det(B^{a_1 \dots a_M}_{i_1\dots i_M})}{\det (B)}
\end{align}
may be computed with Jacobi's theorem as $M\times M$ determinants of the primitives \eqref{eq:primitives} obtained with Cramer's rule.

\subsection{RG: Cluster Amplitudes}
In principle it is possible to extract cluster amplitudes from the cumulants, as they are in one-to-one correspondence. Directly, this means writing the wavefunction \eqref{eq:RGCC2} as the action of an exponential acting upon HF:
\begin{align}
\frac{1}{C_{HF}} \ket{\text{RG}} = \exp \left( \hat{T}_1 + \hat{T}_2 + \dots \hat{T}_M \right) \ket{\text{HF}}
\end{align}
where $\hat{T}_k$ represents a $k$-pair excitation. In practice, this is incredibly tedious, and not particularly informative since the cumulant expressions are so simple. The first cluster amplitudes coincide with the first cumulant
\begin{align}
\hat{T}_1 = \hat{C}_1.
\end{align}
The second-order cluster amplitudes are defined:
\begin{align} \label{eq:T2}
\hat{T}_2 &= \hat{C}_2 - \frac{1}{2} \hat{T}_1 \hat{T}_1
\end{align}
and using the results for $\hat{C}_2$ and $\hat{T}_1$ terms can be collected
\begin{align}
\hat{T}_2 = \sum_{\substack{i,j \in occ \\ a,b \in virt  }} \left( \frac{X^{ab}_{ij}}{Y^{ab}_{ij}} - \frac{X^a_i}{Y^a_i}\frac{X^b_j}{Y^b_j} - \frac{X^b_i}{Y^b_i}\frac{X^a_j}{Y^a_j} \right) S^+_aS^+_b S^-_j S^-_i.
\end{align}
This can be further simplified using \eqref{eq:reduction} and the explicit determinant of a Cauchy matrix:
\begin{align}
\underset{i,\alpha}{\det}\left( \frac{1}{u_{\alpha}- \varepsilon_i} \right) =
\frac{\prod_{a<b} (u_a-u_b)(\varepsilon_b-\varepsilon_a)}{\prod_{ab} (u_a - \varepsilon_b)}
\end{align}
Eventually one arrives at
\begin{align} \label{eq:T2_final}
\hat{T}_2 &= \sum_{\substack{i,j \in occ \\ a,b \in virt  }} \frac{\varepsilon_i - \varepsilon_j}{\varepsilon_a - \varepsilon_b}
\prod_{\substack{m \in occ\\ m \neq i,j}} \frac{(\varepsilon_i - \varepsilon_m)(\varepsilon_j - \varepsilon_m)}{(\varepsilon_a-\varepsilon_m)(\varepsilon_b - \varepsilon_m)} \prod_{\alpha} \frac{(\varepsilon_a - u_{\alpha})(\varepsilon_b - u_{\alpha})}{(\varepsilon_i - u_{\alpha})(\varepsilon_j - u_{\alpha})} \times \nonumber \\
& \left( \frac{(\varepsilon_i - \varepsilon_a)(\varepsilon_j - \varepsilon_b)}{(\varepsilon_i - \varepsilon_b)(\varepsilon_j-\varepsilon_a)} X^a_i X^b_j - \frac{(\varepsilon_i - \varepsilon_b)(\varepsilon_j-\varepsilon_a)}{(\varepsilon_i-\varepsilon_a)(\varepsilon_j-\varepsilon_b)} X^b_i X^a_j  \right) S^+_aS^+_b S^-_j S^-_i
\end{align} 
where the final term in the product is a $2\times 2$ determinant. 

A similar procedure can be followed for
\begin{align}
\hat{T}_3 = \hat{C}_3 - \hat{T}_1\hat{T}_2 -\frac{1}{3!} \hat{T}_1\hat{T}_1\hat{T}_1
\end{align}
and collecting terms yields a structured result:
\begin{align} \label{eq:T3}
\hat{T}_3 = \sum_{\substack{i,j,k \in occ \\ a,b,c \in virt}}
&\left( \frac{X^{abc}_{ijk}}{Y^{abc}_{ijk}} - \mu \left( \frac{X^{ab}_{ij}}{Y^{ab}_{ij}}\frac{X^c_k}{Y^c_k}\right)
+2 \mu  \left( \frac{X^{a}_{i}}{Y^{a}_{i}}\frac{X^{b}_{j}}{Y^{b}_{j}}\frac{X^c_k}{Y^c_k}\right) \right) \times \nonumber \\
& S^+_aS^+_bS^+_c S^-_k S^-_j S^-_i
\end{align}
where we have employed a combinatorial function $\mu$ which generates a sum of all unique \emph{combinations} of the indices of its argument. In the final term of \eqref{eq:T3}, the action of $\mu$ yields the sum of the six unique combinations of indices, which in particular is the permanent of the matrix with entries $\frac{X^a_i}{Y^a_i}$. The second term is a sum of the nine unique combinations: there are 3 choices for the two upper indices and 3 choices for the lower indices. One could also write the first terms as $\mu \left( \frac{X^{abc}_{ijk}}{Y^{abc}_{ijk}}\right)$ since there is only one choice. Equation \eqref{eq:T3} may be simplified to a form similar to \eqref{eq:T2_final}, with an analogous prefactor, but with the $2 \times 2$ determinant replaced with a sum of four $3 \times 3$ determinants.

For $\hat{T}_4$, we will only note briefly that the structure is
\begin{align}
\hat{T}_4 = \sum_{\substack{ i,j,k,l \in occ \\ a,b,c,d \in virt}} &\left( \mu \left( \frac{X^{abcd}_{ijkl}}{Y^{abcd}_{ijkl}} \right) - \mu \left( \frac{X^{abc}_{ijk}}{Y^{abc}_{ijk}}\frac{X^d_l}{Y^d_l} \right) 
- \mu \left( \frac{X^{ab}_{ij}}{Y^{ab}_{ij}}\frac{X^{cd}_{kl}}{Y^{cd}_{kl}} \right)
+2 \mu \left(\frac{X^{ab}_{ij}}{Y^{ab}_{ij}}\frac{X^c_k}{Y^c_k}\frac{X^d_l}{Y^d_l} \right) \right. \nonumber \\
&\left.  
-6 \mu \left( \frac{X^a_i}{Y^a_i}\frac{X^b_j}{Y^b_j}\frac{X^c_k}{Y^c_k}\frac{X^d_l}{Y^d_l} \right) \right)
S^+_aS^+_bS^+_cS^+_d S^-_lS^-_kS^-_jS^-_i
\end{align}
so that a pattern is established. The cluster operators are expressible as $\mu$ acting on the possible partitions of the indices. There is certainly a pattern to the coefficients accessible through symmetric group characters, though we do not consider it useful to deduce. The expressions for the cumulants are quite clean.

\section{Systematic approximations to APIG}
The structure of the cumulants of the RG wavefunction \eqref{eq:big_cumulant} suggests a series of approximations to the APIG coefficients in equation \eqref{eq:APIG_det}. The APIG coefficients are permanents, and thus are intractable to compute in general. We can however, approximate them on an excitation by excitation basis. It is of course understood that this viewpoint supposes that HF is a reasonable first approximation to APIG. While this may be the case for chemical systems, it is certainly not the case in systems that are very strongly-correlated.

The coefficients for the single pair excitations are $1 \times 1$ permanents, i.e. scalars, thus the best first approximation to APIG is AP1roG / pCCD \eqref{eq:ap1rog}. At the level of single pair excitations there is no approximation as the number of variables in each is identical.

Higher pair excitation APIG coefficients require the computation of permanents, which is in general intractable. Rather, we would prefer to compute determinants as their computation scales with the cube of the size of the matrix. The strongest known result in this regard is Muir's theorem \cite{muir:1897}, for any two square matrices $A$ and $B$ of the same size:
\begin{align}
\det(A) \text{per} (B) = \sum_{\sigma} \det (A*B_{\sigma})
\end{align}
where the summation is performed over the $N!$ permutations of the symmetric group and $B_{\sigma}$ is the matrix $B$ with its columns permuted by the action of $\sigma$. Borchardt's theorem is a special case when $A$ is a Cauchy matrix, and $A=B$. Slavnov's theorem is another special case where $A$ is a Cauchy matrix, but the matrix $B$ depends on parameters satisfying Richardson's equations. Muir's theorem is not a tractable recipe on its own, as $N!$ determinants would be required. For $N=2$, there are only two elements of the symmetric group. If we further choose $A=B$, then the permutation switching the two columns on the RHS will give a determinant with identical columns which must of course vanish. Thus, for any $2 \times 2$ matrix $A$
\begin{align}
\text{per} (A) = \frac{\det(A*A)}{\det(A)}.
\end{align}
Thus it is possible to write the double pair excitation APIG coefficients exactly as a ratio of two $2 \times 2$ determinants.

We thus define a wavefunction ansatz which we'll name \emph{pair-coupled cluster determinant ratio} or pCCdr in terms of its cumulants:
\begin{align}
\ket{\text{pCCdr}} &= \sum_k \hat{C}_k \ket{\text{HF}} \\
\hat{C}_k &= \left( \frac{1}{k!}\right)^2 \sum_{\substack{i_1 \dots i_k \in occ \\ a_1\dots a_k \in virt}} 
\frac{\det(A^{a_1\dots a_k}_{i_1\dots i_k} )}{\det(B^{a_1\dots a_k}_{i_1 \dots i_k})} S^+_{a_1}\dots S^+_{a_k} S^-_{i_k} \dots S^-_{i_1}.
\end{align}
where it is understood that $\hat{C}_0 \equiv 1$. Each of the matrices $A^{a_1\dots a_k}_{i_1\dots i_k}$ and $B^{a_1\dots a_k}_{i_1\dots i_k}$ are composed of $k^2$ independent elements. It is possible to express pCCdr as an exponential, though as seen in the previous section this quickly becomes tedious. 

It is easy to see that pCCdrD is pCCD. For pCCdrDQ, there are three independent sets of coefficients $\{t, \alpha, \beta\}$. The cluster operator can be written cleanly in this case, using \eqref{eq:T2}:
\begin{align}
\ket{\text{pCCdrDQ}} &= \exp \left( \hat{T}_1 + \hat{T}_2 \right) \ket{\text{HF}} \\
\hat{T_1} &= \sum_{\substack{i \in occ \\ a \in virt}} t^a_i S^+_a S^-_i \\
\hat{T_2} &= \sum_{\substack{i,j \in occ \\ a,b \in virt}} \left( \frac{\alpha^a_i \alpha^b_j - \alpha^b_i \alpha^a_j}{\beta^a_i \beta^b_j - \beta^b_i \beta^a_j} - t^a_i t^b_j - t^b_i t^a_j \right) S^+_a S^+_b S^-_j S^-_i
\end{align}
Again, up to double pair excitations, this wavefunction is APIG. A naive implementation of pCCdrDQ is more expensive than necessary as the equations for the cluster amplitudes should simplify. We will report on this in an upcoming contribution. Moving to pCCdrDQH, there is now an approximation as a $3 \times 3$ permanent is not equal to a ratio of two determinants. By Muir's theorem, three non-zero contributions would be required. This becomes worse moving to higher excitations. However, in such scenarios HF is likely not a good starting point anyway and RG would itself be a better first approximation.

It should be noted that Zhao and Neuscamman \cite{zhao:2016} have approximated the $2 \times 2$ permanents (APIG double pair excitations) as a a determinant with positive results. Ayers and co-authors \cite{david:2020} have also recently considered projected wavefunctions with ratios of determinants. 

\section{Conclusion}
In this work we have obtained the cumulants of the Richardson-Gaudin wavefunction expressed in the basis of Hartree-Fock Slater determinants. The cumulants have a particularly simple expression that is computable to any order with primitive elements describing single pair excitations. This structure inspires hierarchical approximations to APIG as pair coupled-cluster wavefunctions with ratios of determinants. 

\begin{acknowledgements}
We acknowledge support from the Natural Sciences and Engineering Research Council of Canada (NSERC) as well as the fonds de recherche du Qu\'ebec - nature et technologies (FRQNT).
\end{acknowledgements}

\section*{Conflict of interest}

The authors declare that they have no conflict of interest.


\bibliography{RG_as_CC}

\bibliographystyle{unsrt}

\end{document}